# Social-**Credit +** : AI Driven Social Media Credit Scoring Platform

Thabassum Aslam∗, Anees Aslam∗

## Abstract

*Social-Credit+ is a visionary AI-powered credit scoring system that leverages publicly available social media data {text, images, networks} to augment traditional credit evaluation. It uses a conversational banking assistant to gather user consent and fetch public profiles {LinkedIn, Instagram, etc.}. Multimodal feature extractors {NLP transformers, computer vision models, graph neural networks} analyze posts, bios, images, and friend networks to generate a rich behavioral profile. A specialized Sharia-compliance layer flags any non-halal indicators {e.g. alcohol or gambling content} and prohibited financial behavior {interest, usury} based on Islamic ethics. The platform employs a retrieval-augmented generation {RAG} module: an LLM accesses a domain-specific knowledge base {Islamic finance rules, credit policies} to generate clear, text-based explanations for each decision. We describe the end-to-end architecture and data flow {from data ingestion through feature pipelines and model serving}, the models used {transformers for text, CNN/Vision-Transformers for images, GNNs for networks}, and system infrastructure {streaming, vector DB, LLM}. Synthetic scenarios illustrate how social signals translate into credit-score factors {e.g. "User A: high trust due to verified career info and wholesome lifestyle posts"}. This paper emphasizes conceptual novelty, compliance mechanisms, and practical impact, targeting AI researchers, fintech practitioners, ethical banking jurists, and investors.*

## Introduction

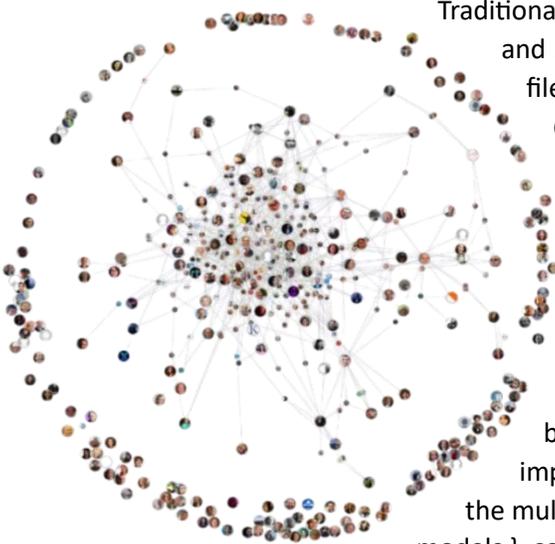

Traditional credit scoring relies on financial history {loans, repayments} and sometimes collateral, which excludes many unbanked or "thin-file" borrowers. By contrast, **Social-Credit+** harnesses alternative data from social media to fill that void. In emerging markets, lenders already eye social profiles to gauge "who you are and if you can be trusted". Our system foregrounds novel data modalities: professional networks, posts, comments, photos, and social graph metrics. It is uniquely tailored to Islamic finance by embedding a Sharia compliance filter: for example, if a user's content repeatedly shows alcohol or gambling, this triggers a non-halal flag. Crucially, all decisions are accompanied by user-friendly explanations generated by a RAG-based chatbot, improving transparency and trust. We detail **Social-Credit+** design: the multi-step data pipeline, the AI models { transformers, GNNs, vision models }, compliance logic, and real-world use cases .

## Related Work

Recent studies show social media can improve credit assessment for those without formal credit histories [1]. For instance, combining traditional financial metrics with social sentiment {e.g. Twitter mood} has enhanced corporate credit ratings[2]. Individual-level work {e.g. using LinkedIn, Weibo} finds features like stable employment or network strength correlate with repayment[2]. Alternative credit scoring systems increasingly fuse diverse data – utilities, rental, social media – to boost inclusivity. Unlike past systems, Social-Credit+ is *multimodal* {text+image+graph} and also integrates generative AI for content analysis and explanation.

Explainable AI {XAI} is critical: opaque models risk untrustworthy decisions. Retrieval-Augmented Generation {RAG} is an emerging XAI approach that combines LLMs with a knowledge base to produce contextual explanations[3]. Guttikonda et al. show a RAG-based chatbot can give "clear, interactive explanations of model outputs" in finance and healthcare[3]. We adopt this idea: Social-Credit+'s RAG module pulls from a Sharia-compliance and credit-policy database to justify approvals or denials in plain language.

In Islamic finance, AI's role is growing. Prior work notes AI can improve the *accuracy and efficiency* of Sharia compliance monitoring[4]. However, Sharia rules impose constraints {no interest, no gambling, no unethical investments} that complicate AI system. Social-Credit+ addresses this by explicitly modeling Sharia ethics: any detected violation {e.g., exposure to riba or explicit haram content} is flagged to ensure adherence.

## System Design

The Social-Credit+ platform consists of the following components:

- **Banking Assistant Bot:** An AI-driven interface that obtains user consent and pulls public social media data via APIs. It ensures privacy by using only user-approved, public content {e.g. LinkedIn profile, Instagram posts}.

- **Text Analysis Module:** Transformer-based NLP models {e.g. BERT, RoBERTa} process biographies, posts, and comments to extract features like employment history, education, sentiment, and language cues. For example, mention of steady job titles or network introductions adds "professional stability" signals, while certain keywords {e.g. "loan", "interest"} might trigger compliance checks.

- **Image Analysis Module:** Computer vision models analyze profile and feed images. We use object-detection/CNN pipelines to classify image content: lifestyles {travel, sports}, assets {cars, homes}, and risk-indicators {presence of alcohol bottles, gambling paraphernalia, fast-food}. Generative vision-language models {e.g. CLIP, image captioners} can describe scenes, enabling analysis of subtle cues {e.g. a beach party vs. a work conference}. Detected features {e.g. partying with liquor} feed into a "lifestyle score" and into the Sharia filte. For instance, if an image contains an **Alcohol** label or a **casino**, an alert is raised.

- **Graph Analysis Module:** A Graph Neural Network {GNN} ingests the user's social graph {friends, connections} to assess network-based trust. Edge weights might represent interaction frequency.

$$h_v^{(k+1)} = \sigma\Big(W \sum_{u \in \mathcal{N}(v)} h_u^{(k)} + b\Big),$$

    Metrics like centrality or clustering further quantify influence or insularity. For example, a highly interconnected professional network {known institutions} could boost a "credibility" feature.

- **Feature Fusion:** Outputs from text, image, and graph modules are concatenated into a unified feature vector. This multi-modal embedding represents the user's social profile. A credit scoring model {e.g. a neural network} takes this vector and computes a creditworthiness score. In abstract terms, one can view the model as combining features:

$$\text{Score} = w_T^\top v_{\text{text}} + w_I^\top v_{\text{image}} + w_G^\top v_{\text{graph}} - \lambda \cdot F_{\text{non-halal}},$$

- **Sharia-Compliance Layer:** This rule-based/ML subsystem flags content or behaviors violating Islamic ethics. It checks for **riba** {interest}, **gharar** {uncertainty}, and **haram lifestyle indicators** {e.g. alcohol, gambling, pornography}. For text, this could be keyword/semantic matching {e.g. mentions of loan interest}. For images, a specialized classifier flags non-halal scenes. Flags then adjust the score downward or trigger manual review. For example, "reading a financial contract for an interest-bearing loan" or "images of alcohol brands" would raise a compliance alert. This aligns with Islamic finance mandates {avoidance of interest and unethical activities}.

- **Explainability via RAG Module:** If a decision {especially a denial} needs justification, an LLM chatbot is invoked. It retrieves relevant policy documents {Sharia guidelines, bank rules} and salient user features from the feature store. For instance, the RAG system might retrieve a bank policy that "active gambling is not acceptable" alongside evidence from the user's data, then generate a natural-language explanation. As Guttikonda et al. demonstrated, a RAG-based XAI can provide "clear, interactive explanations of model outputs" by combining LLM reasoning with domain-knowledge[5].

- **Feedback Interface:** The platform presents a detailed report {via text} to the user, explaining factors like "Job Stability: High {verified career history}", "Spending Patterns: Conservative", "Lifestyle Flag: None", or conversely, any negative flags. This transparency helps meet regulatory and ethical standards.

## Platform Architecture & Data Pipeline

The system's infrastructure integrates streaming ingestion, storage, ML training, Data flows as follows:

1. **Data Ingestion:** The bot collects authorized social data via APIs {e.g. LinkedIn API, Instagram scraping}. This raw data streams into a processing layer {e.g. Apache Kafka or Spark Streaming}.

2. **Preprocessing & Storage:** Raw social text and images are cleaned and stored. Text is tokenized and images resized. Data lakes {e.g. S3 or HDFS} hold the raw corpus, while a feature store {e.g. MongoDB or Cassandra} holds structured features. This design mirrors credit-pipeline patterns: MongoDB's architecture guide shows an ingestion of disparate data into a unified view for modeling.

3. **Feature Extraction:** The NLP and CV models run in batch or real-time to populate the feature store. Graph data is loaded into a graph DB {e.g. Neo4j} for GNN processing.

4. **Model Training:** Training uses GPUs/TPUs. We may deploy pipelines {e.g. Kubeflow} to experiment with different models {transformers for text, CNN/ViT for images, GNNs for graphs}. For example, in practice XGBoost or neural nets have been used, but here we focus on deep models.

5. **Vector Database {for RAG}:** All relevant documents {bank credit policy, Sharia rulings} are encoded into embeddings and stored in a vector DB {e.g. Pinecone, Milvus}. This allows semantic retrieval: given a query {e.g. "User's alcohol image"}, the system fetches related guidelines.

6. **Inference & Serving:** When scoring, the fused features are fed to the credit model {could be in a containerized microservice}. The Sharia-filter runs in parallel. Decisions and supporting facts are logged. If explanation is requested, the RAG service queries the vector DB and generates text using an LLM {e.g. GPT-4 or a fine-tuned domain model}.

7. **System Infrastructure:** The platform leverages cloud services for scalability {compute clusters for training, container clusters for serving}. Data pipelines use Spark or Flink for Message queues ensure reliability. All components have audit logging for compliance and reproducibility.

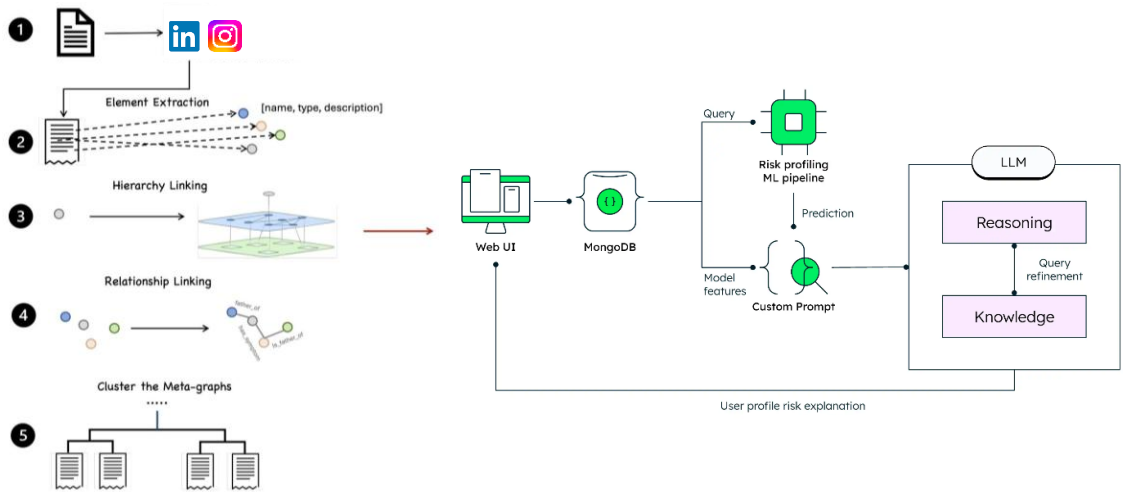

## Compliance Layer

A key innovation is our *Islamic compliance engine*. Islamic finance principles forbid **riba** {interest}, **gharar** {excessive uncertainty}, and **haram activities** {gambling, alcohol, unethical industries}. Social-Credit+ encodes these rules explicitly:

- **Riba/Interest:** The system detects content related to interest-bearing loans or high-risk derivatives. If a user's posts mention "making money from interest" or images of stock trading references, the compliance layer flags it. This aligns with Sharia mandates on avoiding interest.
- **Gambling:** NLP models identify terms like "casino" or "bet," and CV models recognize casinos, poker chips, or lottery cards. If found, the user's score receives a significant penalty. For example, an image of a roulette wheel triggers an immediate red flag.
- **Alcohol and Drugs:** Vision classifiers trained on detecting alcoholic beverages or drug use objects mark any presence of such items. Text analysis likewise catches mentions of drinking or smoking. These occurrences reduce the creditworthiness score, reflecting Islamic ethics on lifestyle.
- **Ethical Investment:** The user's network connections or business interests are scanned for involvement in non-Sharia sectors {e.g. arms, pornography}. For corporate accounts, business descriptions with "interest rates" or "insurance" may trigger scrutiny, since insurance often involves gharar.

This compliance layer provides a binary or graded check: any violation {e.g. one image of alcohol} puts the account in review and reduces the final credit rating. The system ensures transparency by documenting the reason. As noted in Sharia AI research, "maintaining Sharia compliance" complicates AI models, but sophisticated monitoring {AI-driven} can improve consistency. Our design adopts these ideas to ensure that Social-Credit+ respects Islamic norms without human bias.

## Ethics and Privacy

Social-Credit+ is built with ethical safeguards. **Privacy:** We use only user-consented, public data. All data handling complies with privacy laws {GDPR, etc.}. Data is anonymized wherever possible and encrypted at rest. **Transparency:** By using a RAG explanation component, users and regulators can see *why* decisions are made. The system logs an audit trail of features used. **Fairness:** We incorporate bias mitigation techniques {e.g. fairness constraints in training} to avoid discrimination. Because social data can reflect demographics, we ensure decisions cannot exploit protected attributes. For example, if the model flags alcohol-related posts, it does so uniformly, not based on user religion or ethnicity. **Security:** The system follows best practices {secure APIs, access control} to protect sensitive social data. **Human Oversight:** High-risk cases {especially Sharia-violating ones} are reviewed by compliance officers. Such multi-layer checks align with calls for responsible AI in finance.

## Use Cases

- **Financial Inclusion in Emerging Markets:** In regions with low banking penetration, Social-Credit+ enables microloans by using social footprint. For example, a rural entrepreneur with no credit history but an active LinkedIn and successful small business posts can be approved with good terms, enhancing inclusivity. This follows trends where AI opens credit to underbanked groups.

- **Sharia-Compliant Lending:** An Islamic bank uses Social-Credit+ to vet loan applicants. If a Muslim applicant's profile shows frequent interest-based transactions or haram content, the system flags the issue. The lender can then decide to reject or require advisement, ensuring faith-aligned financing. Meanwhile, applicants with clean profiles {charity, community involvement, no alcohol/gambling} receive positive signals.

- **Consumer Loan Decisioning:** A fintech lender integrates our API. During onboarding, the chatbot asks for social profile links. The system quickly analyzes social signals and returns a credit recommendation plus an explanation: "User B was declined because their profile indicates **high-risk behavior**: multiple photos of gambling and no verifiable work history. This violates company policy X." This enhances customer trust by showing data-driven reasons.

- **Credit Education:** Borrowers denied credit can see detailed feedback. For instance, "We noticed frequent party images with alcohol. Reducing such content or adding more professional achievements may improve your score." This educative use-case promotes financial literacy and responsible social media use.

- 

## Synthetic Output

1. **User A – "Professional & Prudent"**:

   - **Data:** LinkedIn shows 10 years at a top engineering firm; Instagram has family photos, travel but no alcohol; network includes verified professionals.

   - **Features:** Strong work history, stable career, no haram indicators; positive social sentiment; moderate spending cues.

   - **Credit Components:** High *Trustworthiness* {due to verified career}, Low *Risk Index*, *Sharia Compliance* = Pass.

   - **Result {Narrative}:** "User A's profile indicates a reliable professional with consistent earnings. Lifestyle images are family-oriented {no alcohol}, and all activities comply with ethical standards. Score: **High**, reasoning: Verified work experience & clean social media."

2. **User B – "Sparse & Risky"**:

   - **Data:** Limited LinkedIn {recent grad}; Instagram shows clubbing with alcohol; network is small {few friends}.

   - **Features:** Thin credit profile, high-risk lifestyle {alcohol detected}, weak network.

- **Credit Components:** Low *Professional Stability*, Elevated *Lifestyle Risk* {alcohol, late-night posts}, *Sharia Compliance* = Fail {alcohol content}.
- **Result:** "User B has an incomplete professional profile and images flagged for alcohol and partying. According to policy, such behavior increases risk. Score: **Low**, due to lack of income stability and a non-halal flag in content."

3. **User C – "Moderate with Compliance Concern"**:
    - **Data:** Successful entrepreneur, but Instagram has one photo of a casino trip; network includes some financial industry contacts.
    - **Features:** Strong business signals, but **Sharia-risk** due to gambling image; network score good.
    - **Credit Components:** *Professional Score* High, *Risk Score* Medium {due to one gambling image}, *Sharia Compliance* = Alert.
    - **Result:** "User C's strong income and business social proof support creditworthiness, but an image of a casino triggered our compliance rules {gambling is prohibited}. Score: **Moderate**, with recommendation: remove such content or clarify context before reassessment."

## Conclusion

Social-Credit+ represents fusion of AI, social data, and ethical compliance for financial services. By combining multimodal analysis {text, image, graph} with Islamic finance rules and explainable AI, it offers a 360° view of creditworthiness that goes beyond traditional scores. Our architecture and pipeline illustrate a practical realization of this vision, addressing data privacy and fairness along the way. In practice, such a system could expand financial access to millions {especially in emerging markets}, while respecting cultural values. Future work includes rigorous fairness audits and live trials. By leveraging generative AI responsibly {RAG explanations, vision models} and embedding Sharia compliance into the model, Social-Credit+ paves the way for more inclusive, transparent, and principled lending in modern era.

## References:

*Alamsyah A., Hafidh A.A., Mulya A.D. {2024}. Innovative Credit Risk Assessment: Leveraging Social Media Data for Inclusive Credit Scoring. J. Risk Financial Manag: [mdpi.commdpi.com](mdpi.commdpi.com).*

*Guttikonda D., Indran D., Narayanan L., Pasarad T., Jha S.B. {2025}. Explainable AI: A RAG-Based Framework for Model Interpretability. In Proc. ICAART 2025 : [scitepress.org](scitepress.org).*

*Shalhoob H., Babiker I. {2025}. Exploration of AI in Ensuring Sharia Compliance in Islamic Finance. Open J. Business & Management, 13{2}, 1435–1448 : [scirp.orgscirp.org](scirp.orgscirp.org).*

*Knowledge@Wharton {2014} "The Surprising Ways Social Media Can Be Used for Credit Scoring." University of Pennsylvania [knowledge.wharton.upenn.eduknowledge](knowledge.wharton.upenn.eduknowledge).*

*SEON {2024}. Social Media Credit Scoring: Pros, Cons, and How to Do It [seon.ioseon.io](seon.ioseon.io).*